\documentclass[10pt]{iopart}
\usepackage{amsfonts}
\usepackage{amssymb}
\usepackage{hyperref}
\usepackage[shortcuts]{extdash}
\usepackage{graphicx}
\usepackage{color}
\usepackage{xcolor}
\usepackage{cite}
\usepackage[mathscr]{euscript}
\usepackage{rotating}
\usepackage{siunitx}
\usepackage{tabularx}

\newcommand{\ud}[1]{{#1^{\dagger}}}
\newcommand{\g}[1]{g^{(#1)}}
\newcommand{\bra}[1]{\left\langle #1\right|}
\newcommand{\ket}[1]{\left| #1\right\rangle}

\newcommand{\mean}[1]{\langle#1\rangle}

\hypersetup{colorlinks=true,linkcolor={{red!80!black}}
  ,citecolor={blue!80!black}, urlcolor={blue!80!black}}

\begin{document} 
\flushbottom 

\title[]{Impact of detuning and dephasing on a laser-corrected
  subnatural-linewidth single-photon source}

\author{J.~C.~{L\'{o}pez~Carre\~{n}o},$^{1, 2}$
  E. Zubizarreta~Casalengua,$^2$ F.~P.~Laussy$^{1,3}$ and E.~del~Valle$^2$} 

\address{$^1$Faculty of Science and
  Engineering, University of Wolverhampton, Wulfruna~St., Wolverhampton
  WV1 1LY, UK}

\address{$^2$Departamento de F\'isica Te\'orica de la Materia
  Condensada, Universidad Aut\'onoma de Madrid, 28049 Madrid, Spain}

\address{$^3$Russian Quantum Center, Novaya 100, 143025 Skolkovo,
  Moscow Region, Russia}
\ead{elena.delvalle.reboul@gmail.com}

\date{\today}

\begin{abstract}
  The elastic scattering peak of a resonantly driven two-level system
  has been argued to provide narrow-linewidth antibunched
  photons. Although independent measurements of spectral width on the
  one hand and antibunching on the other hand do seem to show that
  this is the case, a joint measurement reveals that only one or the
  other of these attributes can be realised in the direct emission. We
  discuss a scheme which interferes the emission with a laser to
  produce simultaneously single photons of subnatural linewidth. In
  particular, we consider the effect of dephasing and of the detuning
  between the driving laser and/or the detector with the emitter. We
  find that also in presence of dephasing, our scheme brings
  considerable improvement as compared to the standard scheme.
\end{abstract}

\section{Introduction}
\label{sec:WedApr18145430BST2018}

Of the many schemes to implement a single-photon
source~\cite{laucht12a,schlehahn18a,krapick13a,zadeh16a,gazzano13a,kim16b,neergaardnielsen07a,portalupi15a,reimer12a,toishi09a,he17a,santori02a,he13a,muller16a,na10a,dada16a,ding16a,somaschi16a,loredo17a,santana17a},
various strategies are adopted to favour one or the other of its
desirable characteristics, ranging from its size and ability to
interface with other optical
devices~\cite{laucht12a,schlehahn18a,krapick13a,zadeh16a}, its
brightness~\cite{gazzano13a, kim16b, neergaardnielsen07a,
  portalupi15a, reimer12a, toishi09a,he17a}, the indistinguishability
between successive photons~\cite{he17a,santori02a, he13a, muller16a,
  na10a,dada16a, ding16a} and, of course, its sub-Poissonian
character~\cite{somaschi16a, loredo17a, santana17a}.  More exotic
schemes have also been proposed, such as the photon blockade, in both
its conventional~\cite{verger06a, ridolfo12a} and
unconventional~\cite{liew10a, gerace14a, flayac15a} versions (both of
which have been recently demonstrated
experimentally~\cite{arXiv_munozmatutano17a,arXiv_delteil18a,arXiv_snijders18a,arXiv_vaneph18a})
or the heralding of single photons from two-level systems driven in
Mollow triplet regime~\cite{lopezcarreno17a}, let alone sources of
$N$-photons ``bundles''~\cite{sanchezmunoz14a, sanchezmunoz18a}.  One
of the most popular platforms for the generation of single photons,
both from an experimental and theoretical point of view, is a
two-level system. This can be realised in a variety of platforms
ranging from cold atoms~\cite{itano88a, grangier86b,rempe90a}, to
semiconductor quantum dots~\cite{michler00a,lounis00b, santori01a,
  zwiller02b, sebald02a,santori02a, pelton02a,yuan02a, gerardot05a}
passing by ions~\cite{diedrich87a,bergquist86a,schubert92a},
molecules~\cite{kask85a,basche92a,treussart01a,lounis00a},
superconducting circuits~\cite{astafiev10a,bozyigit11a, lang13a,
  hoi13a, gu17a}, nitrogen
vacancies~\cite{kurtsiefer00a,brouri00a,messin01b}, among others.  A
priori, a two-level system fits perfectly the purpose, as it can only
sustain a single excitation at any given time. Thus, its repetition
rate is limited by the time it takes to ``reload'', and one can expect
a perfectly antibunched emission.  This is however a simplified
description that ignores a central aspect of quantum theory: the
detection process. The two-level system is characterized to the best
of its abilities only by a detector that can measure its emission with
infinite precision in time. Conversely, if the detector has a finite
temporal resolution (as is of course the case in any actual setup),
or, equivalently, a finite bandwidth, the theoretically perfect
suppression of the second-order correlation function is spoiled by the
Heisenberg uncertainty principle~\cite{arXiv_delvalle17a}. This can be
described accurately by the theory of frequency-filtered
correlations~\cite{delvalle12a}.

While the impact of detection is a fundamental principle that applies
to all quantum systems, an interesting and somehow counter-intuitive
effect occurs when turning to the detected emission of a two-level
system driven coherently in the so-called Heitler
regime~\cite{heitler_book44a}, in which the emission of a two-level
system consists of two components: i) photons that are absorbed and
later re-emitted (fluorescence) and ii) photons that are elastically
scattered by the two-level system (in a coherent absorption and
re-emission process). The former are emitted with a Lorentzian profile
centered at the frequency of the driving laser and with the natural
linewidth of the two-level system, constituting the incoherent
fraction of the emission. The latter are emitted as a $\delta$-narrow
peak (assuming a vanishing linewidth for the laser), which forms the
coherent fraction of the emission that dominates at low driving. Like
any two-level system, the total emission is antibunched. The idea then
arose to use the $\delta$ peak to collect antibunched photons with
narrow spectral width~\cite{matthiesen12a, nguyen11a}. Here as well,
one must not forget the process of detection, and taking it into
account, we have shown that these two qualities are not realised
jointly~\cite{arXiv_lopezcarreno18a}: the detected photons are either
antibunched, but with a spectral width no better than that of the
emitter itself, or they can be detected with the spectral bandwidth of
the $\delta$ peak, but then their antibunching is dramatically
reduced.  Interestingly, however, we have shown in the same
work~\cite{arXiv_lopezcarreno18a} how to detect photons jointly with a
subnatural linewidth and an excellent anibunching, by interfering the
emission of the filtered two-level system with an external laser. This
laser correction removes, through destructive interferences, the
excess of coherent emission when focusing on the $\delta$ peak, in a
process akin to an homodyne interference~\cite{vogel95a,
  collett87a}. Similar schemes have been recently implemented to
obtain a source of indistinguishable photons~\cite{muller16a}, to
observe the rising of the so-called dynamical Mollow
triplet~\cite{fischer16a} and to unveil the photon correlations of the
light emitted by a Jaynes-Cummings system~\cite{fischer17a}. In our
case, we find that not only this laser-correction allows to realise
simultaneously subnatural linewidth spectral emission and
antibunching, but also that it produces a stronger type of
single-photon emission with a plateau in the time-resolved photon
correlation~$g_a^{(2)}(\tau)$. Such sources therefore provide a new
playground of their own, whose properties, advantages over existing
sources and further possibilities deserve an immediate attention, as
we wait for their experimental implementation.

In this text, we provide a more general picture, including other
interesting features of the statistics, such as perfect superbunching
(where, to first order in the driving, $g_a^{(2)}(0)$ becomes infinite)
in addition to the previously reported perfect antibunching
($g_a^{(2)}(0)=0$). More particularly, we focus on the effect of two
important aspects not considered previously: the impact of dephasing,
since this is a detrimental ingredient that is typically present,
especially in a solid-state platform, and the role of detuning from
the two-level system, from either the driving laser and/or the
detector.

\section{Theoretical description}
\label{eq:TueMar13133354GMT2018}

We consider a two-level system driven by a coherent source in the
regime of low excitation, commonly referred to as the Heitler regime,
with Hamiltonian (we take~$\hbar=1$ along the text)
\begin{equation}
  \label{eq:2LShamiltonian}
  H_{\sigma} = (\omega_\sigma-\omega_\mathrm{L}) \ud{\sigma} \sigma +
  \Omega_{\sigma} 
  \left(\ud{\sigma} + \sigma \right)\,.
\end{equation}
The two-level system has a frequency~$\omega_\sigma$ and is described
through an annihilation operator~$\sigma$ that follows the pseudo-spin
algebra, whereas the laser is treated classically, i.e., as a
$c$-number, with intensity~$\Omega_\sigma$ and
frequency~$\omega_\mathrm{L}$. The dissipative character of the system
is included in the dynamics through a master equation
\begin{equation}
\label{eq:2LSmasterEq}
\partial_t \rho = i \left[ \rho ,H_{\sigma}  \right] + \frac{\gamma_\sigma}{2}
\mathcal{L}_{\sigma} \rho + \frac{\gamma_\phi}{2}
\mathcal{L}_{\ud{\sigma}\sigma} \rho\,,
\end{equation}
where~$\gamma_\sigma$ is the decay rate of the two-level system,
$\gamma_\phi$~is the dephasing rate, $H_\sigma$ is the Hamiltonian in
Eq.~(\ref{eq:2LShamiltonian}) and
$\mathcal{L}_{\sigma} \rho = 2 \sigma \rho \ud{\sigma} - \ud{\sigma}
\sigma \rho - \rho \ud{\sigma} \sigma$.  The steady-state solution of
Eq.~(\ref{eq:2LSmasterEq}) can be written in terms of two magnitudes:
the population, $\mean{\ud{\sigma}\sigma}\equiv n_\sigma$, and the
coherence, $\mean{\sigma}\equiv \alpha$, of the two level system:
\begin{equation}
\label{eq:2LSrhoSS}
\rho_{\mathrm{ss}} = \left (
\matrix{
1 - n_\sigma & \alpha \cr
\alpha^\ast & n_\sigma\cr
} \right)\,,
\end{equation}
where
\numparts
\begin{eqnarray}
  \label{eq:TueMar13135716GMT2018}
  n_\sigma &=& \frac{4(\gamma_\sigma+\gamma_\phi)
    \Omega_{\sigma}^2}{\gamma_\sigma[(\gamma_\sigma+\gamma_\phi)^2 + 4 
  \Delta_{\sigma}^2] + 8 (\gamma_\sigma+\gamma_\phi)
  \Omega_{\sigma}^2}\,,\\
\alpha &=& \frac{2i \gamma_\sigma \Omega_{\sigma}  [2i \Delta_\sigma - (
  \gamma_\sigma+ \gamma_\phi)]}{\gamma_\sigma[(\gamma_\sigma+\gamma_\phi)^2
           + 4  
  \Delta_{\sigma}^2] + 8 (\gamma_\sigma+\gamma_\phi)
  \Omega_{\sigma}^2}\,,
\end{eqnarray}
\endnumparts
with~$\Delta_\sigma=(\omega_\sigma-\omega_\mathrm{L})$ the detuning
between the two-level system and the driving laser.

To model the detection process self-consistently, one can couple to
this system a detector and study the observables through this detector
rather than from the object itself. A method (the so-called ``cascaded
formalism'') has been developed in the early
90s~\cite{gardiner93a,carmichael93b} to model this theoretically. Such
a precaution avoids (or reveals) the subtle mistake of assuming that
the emitted light retains all the attributes of the source when these
are computed or measured separately. To take the example of interest
in our discussion, one can indeed observe (or compute) the spectral
width of the two-level system in the Heitler regime, and find an
ultra-bright and ultra-narrow component, and then observe (or compute)
its antibunching and find an excellent antibunching. These constitute
separate characterisations of the source, and until these are measured
simultaneously, they cannot be assumed to exist
simultaneously. Indeed, characterising the light through the
detector---which by the very nature of its excitation is being subject
to the both aspects of interest simultaneously---one finds that the
detected light is either antibunched but not narrower than the source,
or is spectrally narrow but then with a poor or no antibunching. This
depends on the spectral width of the detector itself: if the detector
has a large spectral width, it will not be sensitive to the supposedly
narrow linewidth light that excites it. If the detector has a narrow
spectral width, it will not be sensitive to its antibunching. Given
the character of quantum mechanics, we conclude that the failure of a
detector to simultaneously capture the narrow linewidth and the
antibunching really means that these do not actually jointly exist.

In the following we will discuss and generalise a scheme which we have
recently proposed~\cite{arXiv_lopezcarreno18a} and that achieves such
a joint narrow and antibunched emission, in the sense that a detector
does collect its light with these two attributes intact. As we will
focus on antibunching and spectral width, we can support our analysis
of the detection process through a ``sensor'' that acts as a filter
for the emitted light~\cite{delvalle12a}. Theoretically, this is
included through the vanishing coupling of a bosonic field with
annihilation operator~$a$ to the dynamics of the two-level system, by
adding the Hamiltonian $H_a=g(\ud{\sigma}a+\ud{a}\sigma)$ to
Eq.~(\ref{eq:2LShamiltonian}) and then taking the limit
$g\rightarrow0$, which allows the dynamics of the two-level system to
be independent from that of the sensor. The bandwidth of the sensor is
given by its decay rate~$\Gamma$ and is included as an extra
term~$(\Gamma/2)\mathcal{L}_a\rho$ in the master
equation~(\ref{eq:2LSmasterEq}). For quantities such as populations,
which would vanish with~$g\rightarrow0$, one should use instead the
more complete but also heavier cascaded formalism, which we have shown
is equivalent to the more lightweight sensor method as far as
correlations are concerned~\cite{lopezcarreno18a}. The main point of
this theoretical shortcut is that instead of considering the light
emitted by a system, one can consider instead the filtered light and
this is enough to describe the process of detection, as long as the
detector would have the same spectral width as the filter. We will
therefore be speaking of filtering for the light emitted by the
two-level system, which should be understood as the effect of its
detection from a detector with the corresponding bandwidth.

When the emission of the two-level system is filtered in frequency,
the tails of the incoherent Lorentzian are trimmed out and this spoils
the perfect antibunching, that arises from an interference between the
coherent and incoherent components~\cite{arXiv_lopezcarreno18a}.  Such
an interference can be restored simply by reinstating the original
proportion, i.e., the perfect antibunching can be maintained after
filtering by reducing the surplus of coherent emission that passes in
its entirety through the filter. This can be easily achieved since a
coherent field can be scaled at will through interferences, in our
case, destructive interferences. Our scheme thus consists of
interfering at a beam splitter the light emitted by the two-level
system with a coherent field $\beta \equiv |\beta|e^{i\phi}$, whose
amplitude and phase need to be fixed adequately to provide the exact
compensation. In this configuration, the Hamiltonian for our
laser-corrected source becomes
\begin{equation}
  \label{eq:TueMar13171915GMT2018}
  H= H_\sigma + \Delta_a \ud{a}a - i r |\beta|(e^{i\phi} \ud{a}
  - e^{-i\phi}a)  + g t (\ud{\sigma}a + \ud{a}\sigma)\,,
\end{equation}
where~$H_\sigma$ is the Hamiltonian in Eq.~(\ref{eq:2LShamiltonian}),
$t$ and~$r$ are the transmission and reflection coefficients of the
beam splitter (the reflection coefficient is preceded by a factor~$i$,
which accounts for the phase shift gained by the reflection in the
beam splitter), and~$\Delta_a=(\omega_a - \omega_\mathrm{L})$ is the
detuning between the detector and the driving laser. The amplitude of
the coherent field~$|\beta|$ can be parameterised as a
fraction~$\mathcal{F}$ of the coherent field that the sensor receives
from the two-level system, i.e., we may write
\begin{equation}
  \label{eq:SatFeb24130524CET2018}
  |\beta| =
  g\frac{\Omega_\sigma}{\gamma_\sigma}\frac{t}{r}\mathcal{F}\,, \quad
  \quad \mathrm{or\,\,\,equivalently} \quad \quad
  |\beta'|= g\frac{\Omega_\sigma}{\gamma_\sigma}\mathcal{F}\,,
\end{equation}
where we have also defined~$|\beta'|=|\beta|(r/t)$, to avoid
carrying the parameters of the beam splitter, which only renormalize
the observables but do not change the physics involved.

\section{Results}
\label{sec:FriApr20161034BST2018}

The two-photon correlations detected in various spectral widths are
obtained from the steady-state solution of the master
equation~(\ref{eq:2LSmasterEq}) with $H_\sigma$ replaced by the
Hamiltonian in Eq.~(\ref{eq:TueMar13171915GMT2018}) and with the added
Lindblad term~$(\Gamma/2)\mathcal{L}_a\rho$.  Although these
correlations can be obtained in closed-form, they are too cumbersome
to be written here. Instead, we will provide the particular cases
i)~with detuning but no dephasing and ii)~at resonance but with
dephasing. The full case, of which we will show one case graphically,
brings little more insights, so treating these separately is enough to
reach general conclusions. We consider first the case without
dephasing.

\subsection{No dephasing}

In the case of no dephasing, but allowing for some detuning, either
between the driving laser and the two-level system or between the
detector and the two-level system (or both), the detected two-photon
correlations are given by, to leading order in the driving strength:
\begin{equation}\fl
    \label{eq:MonMar12165557GMT2018}
    \begin{array}{r@{}l}
  g_a^{(2)} &= \tilde \Gamma_\sigma^2 \big(16 \gamma_\sigma^4 + 16
  \mathcal{F}^2 \gamma_\sigma^2 \tilde \Gamma_+^2 + \mathcal{F}^4
  \tilde \Gamma_\sigma^2 \tilde \Gamma_+^2 - {}\\ \fl
 & \quad \quad {}-8 \mathcal{F} \gamma_\sigma
  \{ 2 (4\gamma_\sigma^2 \Delta_+ + \mathcal{F}^2 \Delta_\sigma
  \tilde \Gamma_+ ) \sin \phi+
  2 \gamma_\sigma \mathcal{F} (\gamma_+ \Delta_\sigma + \gamma_\sigma
  \Delta_+)\sin
  2\phi-{}\\ \fl
  &\quad \quad
  {}-\gamma_\sigma (4\gamma_\sigma \gamma_+ + \mathcal{F}^2
   \tilde \Gamma_+^2) \cos \phi   -
  \gamma_\sigma \mathcal{F} (\gamma_\sigma \gamma_+ - 4\Delta_\sigma 
  \Delta_+) \cos 2\phi \} \big) \big / {}\\ \fl
  &\quad \quad
  \big ( \tilde \Gamma_+^2 [4 \gamma_\sigma + \mathcal{F}^2 \tilde\Gamma_\sigma
+4 \gamma_\sigma\mathcal{F}(
  \gamma_\sigma \cos\phi -2 \Delta_\sigma^2\sin \phi)]^2  \big)\,,
    \end{array}
\end{equation}
where we have defined~$\gamma_+ \equiv \gamma_\sigma + \Gamma$,
$\Delta_+ \equiv \Delta_\sigma + \Delta_a$
and~$\tilde \Gamma_c^2 \equiv \gamma_c^2 + 4\Delta_c^2$
for~$c=\sigma,+$. From this expression, it is easy to find particular
cases of interest: perfect antibunching, when the numerator becomes
zero, and perfect superbunching, when the denominator becomes zero.

The condition to vanish the numerator, and therefore to produce
perfect antibunching, is given by
\begin{equation}
  \label{eq:MonMar12172809GMT2018}
  \mathcal{F} =  -\frac{2\gamma_\sigma e^{-i\phi}}{\gamma_\sigma +2
    i\Delta_\sigma} \left ( 1 \pm \sqrt{
      \frac{\Gamma+2i\Delta_a}{\Gamma_+ +2i\Delta_+} } \right)\,,
\end{equation}
which generalizes the expression given in Eq.~(10) of
Ref.~\cite{arXiv_lopezcarreno18a}. Since a real phase~$\phi$ can
always be found so that Eq.~(\ref{eq:MonMar12172809GMT2018}) is a real
positive number, the condition for perfect antibunching is always
satisfied. The condition that cancels
Eq.~(\ref{eq:MonMar12165557GMT2018}) also yields the suppression of
the state with two photons in the detector, as we show in~\ref{app:3}
through a wave-function approximation.
\begin{figure*}
  \centering
  \includegraphics[width=\linewidth]{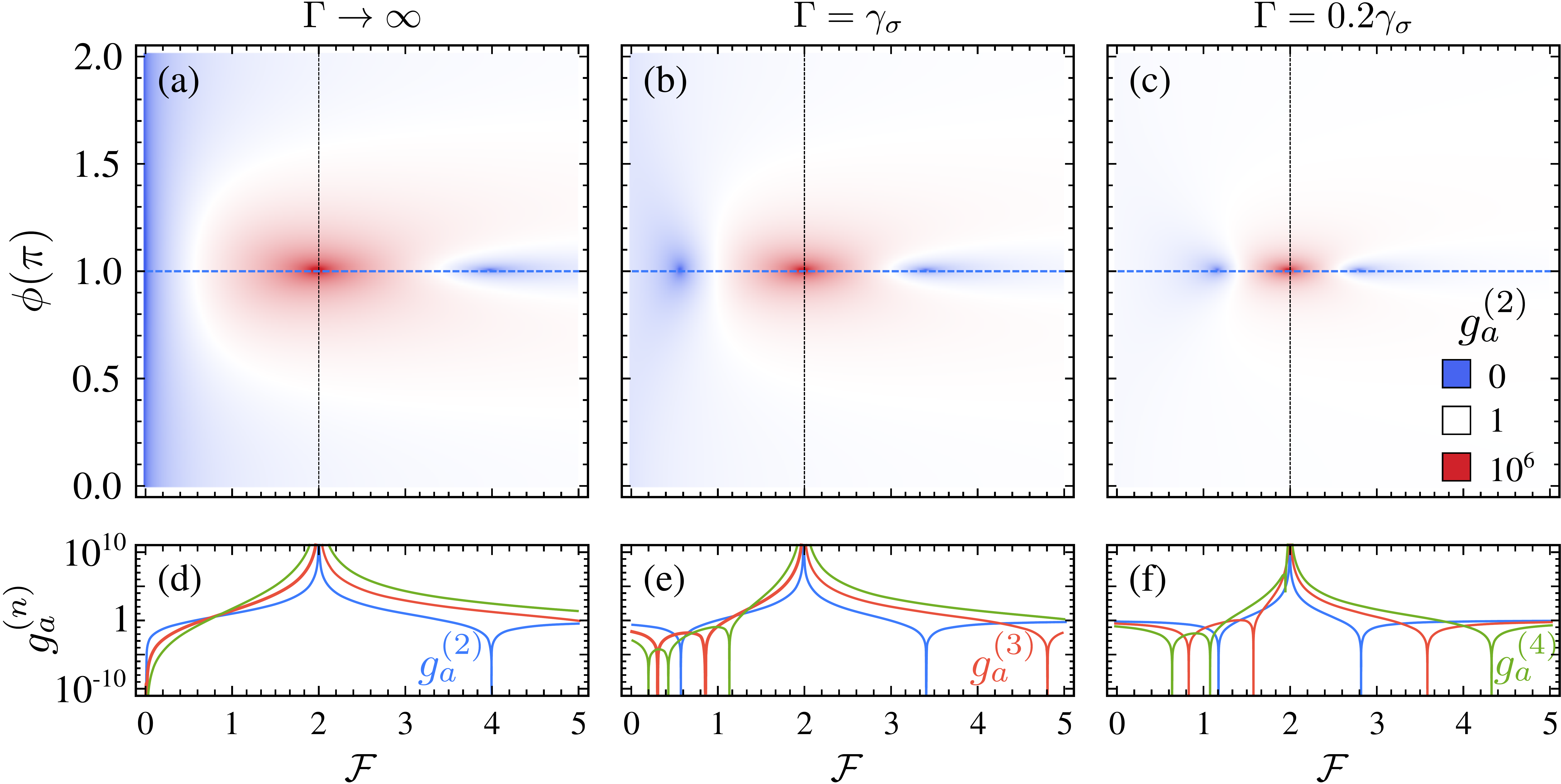}
  \caption{(Color online)~Filtered two-photon correlations at
    resonance and without dephasing, as a function of the
    parameters~$\mathcal{F}$ (intensity) and~$\phi$ (phase) of a
    superimposed interfering laser. In all the panels we
    set~$\gamma_\sigma$ as the unit, $\Delta_\sigma=\Delta_a=0$ and
    $\gamma_\phi=0$. The detector linewidth decreases in from (a) to
    (c) as indicated on each panel.}
  \label{fig:WedNov29150123CET2017}
\end{figure*}
In the most natural configuration where the laser is resonant to both
the cavity and the two-level system,
Eq.~(\ref{eq:MonMar12172809GMT2018}) reduces to the expressions of
Ref.~\cite{arXiv_lopezcarreno18a}:
\begin{equation}
  \label{eq:WedMar14161430GMT2018}
  \mathcal{F}_{2,\pm} = 2 \left(1 \pm
    \sqrt{\frac{\Gamma}{\Gamma+\gamma_\sigma}} \right) \quad \mathrm{and}
  \quad \phi_+ = \pi\,.
\end{equation}

On the other hand, the denominator of
Eq.~(\ref{eq:MonMar12165557GMT2018}) vanishes when one sets the
phase~$\phi$ and amplitude~$\mathcal{F}$ of the interfering laser to
satisfy
\begin{equation}
  \label{eq:MonMar12171956GMT2018}
    \tan \phi = - \frac{2\Delta_\sigma}{\gamma_\sigma} \quad \quad
    \mathrm{and} \quad\quad
    \mathcal{F}= -2 \cos \phi\,.
\end{equation}
Note that unlike for antibunching, this condition to obtain
superbunching is independent of the detuning between the detected
light and the driving laser. In this case, $g_a^{(2)}$ diverges (to
first order in the driving strength, so that higher order terms would
produce huge but finite values of~$g_a^{(2)}(0)$).  The appearance of
such a strong superbunching from a two-level system admixed with a
coherent field is interesting. Its nature is discussed
elsewhere~\cite{arXiv_zubizarretacasalengua18a}. Here, we will content
to say that this arises from an interference, as a result of which the
photon emission is not occurring in the form of photon bundles and
cannot be Purcell-enhanced, so that prospects for applications as~$N$
photon sources are limited. Also, we mention that infinite bunching
has already been reported before; it occurs for instance with Fock
states in bosonic cascades~\cite{liew16a}.

A full map of~$g_a^{(2)}(0)$ as defined by
Eq.~(\ref{eq:MonMar12165557GMT2018}) is shown in
Figure~\ref{fig:WedNov29150123CET2017}(a-c) for spectral widths of the
detector ranging from essentially full-bandwidth, panel~(a), the
linewidth of the two-level system, panel~(b) and with sub-linewidth
resolution, panel~(c). In the first case, without frequency filtering,
perfect antibunching is obtained without any laser correction, that
is, for~$\mathcal{F}=0$ and independently of the phase. This
corresponds to the case considered in the
literature~\cite{matthiesen12a, nguyen11a}, but this comes at the cost
of the spectral width: the $\delta$-width of the laser is completely
washed out by the detector. For a detector spectrally matched to the
emitter, shown in~(b), antibunching is considerably reduced by the
detector (to~$g_a^{(2)}(0)\approx 0.25$). Keeping the same linewidth,
antibunching can be restored by the interfering laser fulfilling
conditions~(\ref{eq:WedMar14161430GMT2018}), restoring an exact
antibunching, $g_a^{(2)}(0)=0$, to first order in the driving. Going
to sub-natural linewidth with a detector spectrally matched
to~$0.2\gamma_\sigma$, shown in panel~(c), one finds that antibunching
is now almost completely gone in absence of the laser correction,
$g_a^{(2)}(0)\approx 0.7$, but can again be fully restored with the
laser correction. In all cases, in between the two conditions for
antibunching, one can see the superbunching at~$\mathcal{F}=2$.  Note
that, as the linewidth is made narrower, the conditions for
antibunching, Eq~(\ref{eq:WedMar14161430GMT2018}), come closer to the
conditions for superbunching, Eq.~(\ref{eq:MonMar12171956GMT2018}). So
while this effect could be pursued down to extremely narrow
linewidths, at no cost for the antibunching, some restrictions would
arise from the stability of the driving laser, as energy fluctuations
would take the system from the condition for perfect antibunching to
the condition for superbunching. Panels~(d-f) show in blue lines
transverse cuts at~$\phi=\pi$ of panels~(a-c), and in red and green
lines the corresponding higher-order correlators~$g^{(3)}_a$
and~$g_a^{(4)}$, respectively. This shows how, although
both~$\mathcal{F}_{2,\pm}$ from Eq.~(\ref{eq:WedMar14161430GMT2018})
yield an exact cancellation of~$g_a^{(2)}$, the higher order
correlations remain sub-Poissonian in the vicinity of
only~$\mathcal{F}_{2,-}$ (which corresponds to the condition to obtain
a ``conventional
antibunching''~\cite{arXiv_zubizarretacasalengua18a}).

\subsection{Dephasing}

We now turn to the impact of dephasing, which is detrimental to photon
correlations, but can still be corrected to a considerable extent
through our process, although not perfectly anymore. In the case of
dephasing alone, where both the two-level system and the detector are
resonant with the driving laser, the two-photon correlations are given
by
\begin{equation}\fl
  \label{eq:WedApr18101223BST2018}
  \begin{array}{r@{}l}
    g_a^{(2)} &= \Gamma_\phi (\Gamma+\Gamma_\phi) \big ( 16
                \gamma_\sigma^3 (\Gamma+\gamma_+) (2\Gamma+\gamma_+) +16
                \mathcal{F}^2 \gamma_\sigma \gamma_+^2 (2\Gamma +
                \Gamma_\phi) (3\Gamma+ \Gamma_\phi)+\\ \fl
              & \quad  {}+\mathcal{F}^4 \gamma_+ \Gamma_\phi
                (\Gamma+\Gamma_\phi) (2\Gamma+\Gamma_\phi) (3\Gamma+\Gamma_\phi)+{}\\ \fl
              &\quad
                {}+8 \mathcal{F}\gamma_\sigma \gamma_+ (3 \Gamma+\Gamma_\phi) \{
                \mathcal{F} \gamma_\sigma (2\Gamma+\Gamma_\phi) \cos 2\phi +{}\\ \fl
              &\quad
                {}+ [4\gamma_\sigma (\Gamma+\gamma_+) + \mathcal{F}^2 (\Gamma +
                \Gamma_\phi) (2\Gamma+\Gamma_\phi)] \cos \phi \}  \big) \big / {}\\ \fl
              &\quad
                \big \{  \gamma_+^2 (2\Gamma + \Gamma_\phi) (3\Gamma +
                \Gamma_\phi) \times {}\\ \fl 
              &\quad \quad \quad \quad \quad \quad
                {} [(4+\mathcal{F}^2)\gamma_\sigma \gamma_+ + \mathcal{F}^2 \gamma_\phi (\gamma_+ + \Gamma_\phi)+4
                \mathcal{F} \gamma_\sigma (\Gamma + \Gamma_\phi)\cos \phi]^2  \big\}\,,
  \end{array}
\end{equation}
where we have used the
notation~$\Gamma_\phi = \gamma_\sigma+\gamma_\phi$
and~$\gamma_+ = \gamma_\sigma+\Gamma$. 

\begin{figure}
  \centering
  \includegraphics[width=\linewidth]{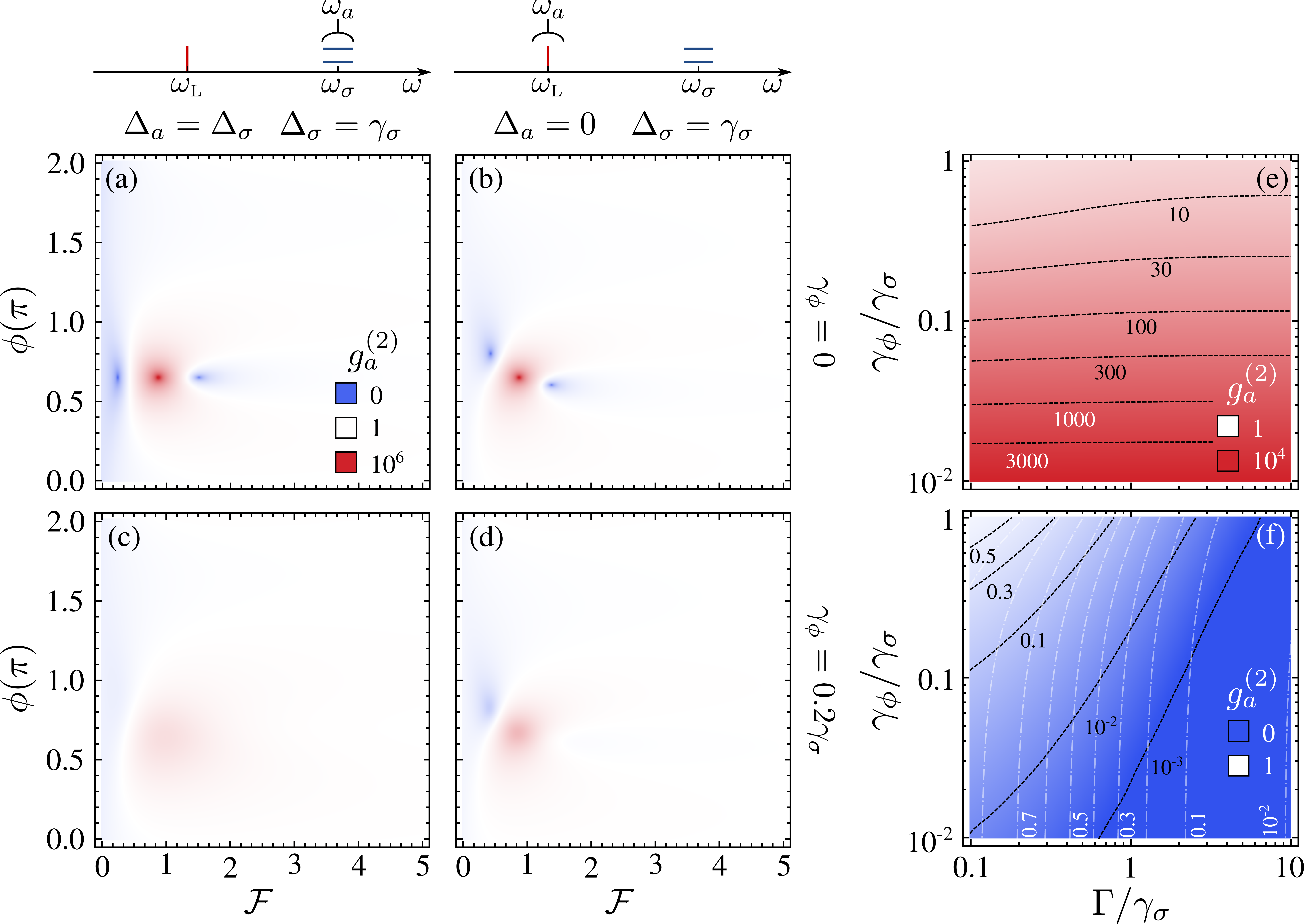}
  \caption{(Color online) (a--d) Two-photon correlations as a function
    of the parameters of the interfering laser when the laser is
    detuned from the two-level system, in absence (top row) and in
    presence (bottom row) of dephasing. In the left column, the
    detector is resonant to the emitter while in the right column, it
    is to the driving laser. Dephasing spoils the perfect
    antibunching, and it is particularly detrimental in the case where
    the cavity is set in resonance the two-level system, in which case
    the dephased correlations are completely blurred. Without
    dephasing, one may still find the condition for~$\mathcal{F}$ that
    provides perfect antibunching. (e,f)~Maximum and minimum value
    of~$g^{(2)}_a$ when the two-level system is dephased with
    $\gamma_\phi=0.2\gamma_\sigma$. In panel~(f) we also show in
    dashed-dotted white lines the isolines for the minimum
    correlations that can be obtained without the correction (i.e.,
    with~$\mathcal{F}=0$). Throughout, $\Gamma=\gamma_\sigma/5$.}
  \label{fig:TueMar13115744GMT2018}
\end{figure}

The general case that also includes detuning is shown in
Fig.~\ref{fig:TueMar13115744GMT2018}.  In this case, the laser is
detuned from the two-level system, while the detector is set either at
the frequency of the two-level system (panels~(a) and~(c)) or at the
frequency of the laser (panels~(b) and~(d)). The top row of
Fig.~(\ref{fig:TueMar13115744GMT2018}) is given by
Eq.~(\ref{eq:MonMar12165557GMT2018}) while the bottom row, for which a
closed-form expression exists but is to bulky to be written here, is
only shown graphically.  The filter linewidth, or, equivalently, the
bandwidth of the detector, has been taken to match one fifth of the
emitter linewidth. Therefore, Fig.~\ref{fig:TueMar13115744GMT2018}(a)
and Fig.~\ref{fig:TueMar13115744GMT2018}(b) are the detuned versions
of Fig.~\ref{fig:WedNov29150123CET2017}(c). Using this panel as a
reference, one can see the impact of dephasing (spoiling the
correlations) and detuning (maintained but for different laser
corrections). Actually, the condition for superbunching is independent
from the detuning between the detector and the driving laser, in
agreement with Eq.~(\ref{eq:MonMar12171956GMT2018}), unlike the
condition for antibunching. In fact, when the detector is resonant to
the laser, both being detuned from the emitter, the conditions for
perfect antibunching do not occur at the same phase~$\phi$. Therefore,
in a detuned measurement, the correlations are easier to observe when
the cavity is resonant to the two-level system.

While in absence of dephasing (Eq.~(\ref{eq:MonMar12165557GMT2018})),
correlations can range from exactly zero to infinity,
Eq.~(\ref{eq:WedApr18101223BST2018}) shows that in its presence, they
can only be pushed to finite values, both for bunching and
antibunching.
Although the full expressions to obtain such limiting case can be
found, they are too bulky to be written here but they are plotted in
Fig.~\ref{fig:TueMar13115744GMT2018}(e). Similarly, the loss of
antibunching due to dephasing cannot be compensated exactly by the
interfering laser and for strong enough dephasing, no antibunching at
all can be maintained. Nevertheless, the laser correction still brings
considerable improvement on the case without interference, which is
also, of course, affected by dephasing. The intensity~$\mathcal{F}$
that yields the best antibunching is found by minimising
Eq.~(\ref{eq:WedApr18101223BST2018}), which does not provide a simple
closed-form expression, but can be readily found numerically. The
minimum antibunching obtained this way is shown in
Fig.~\ref{fig:TueMar13115744GMT2018}(f).

\subsection{Correlations in time}

\begin{figure}
  \centering
  \includegraphics[width=\linewidth]{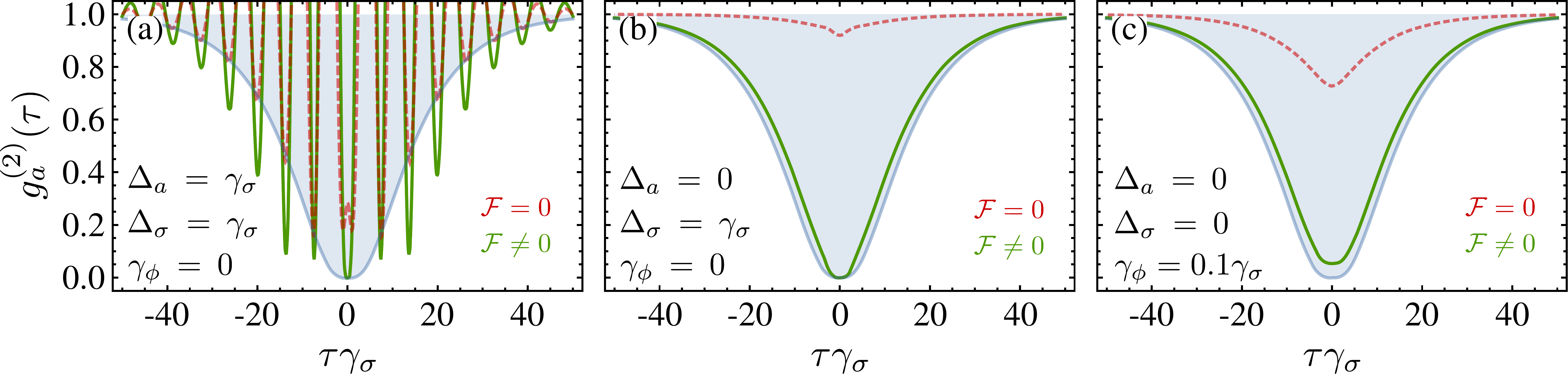}
  \caption{(Color online)~Time-resolved filtered two-photon
    correlations in the various configurations discussed in the
    text. The filled-blue line is the case at resonance and without
    dephasing. Panels~(a) and~(b) show the impact of various detunings
    without dephasing and Panel~(c) shows the impact of dephasing at
    resonance. In green the best correction that can be achieved with
    an interfering laser, in red the uncorrected case. In all cases
    $\Gamma=\gamma_\sigma/5$.}
  \label{fig:MonApr16210113BST2018}
\end{figure}

While the value of the zero-delay correlation between photons is
usually the one considered to quantify the sub-Poissonian character of
a source, the correlations between photons detected with a time
difference~$\tau$ are also important. In particular, fast oscillations
in correlations can be difficult to resolve and average out the
result. In Ref.~\cite{arXiv_lopezcarreno18a}, we showed that in the
case without dephasing and in resonance, when the loss of antibunching
due to filtering is corrected with an external laser,
the~$g_a^{(2)}(\tau)$ displays a plateau of perfectly antibunched
photons for up to~$|\tau|\approx 2.5\gamma_\sigma$. This actually
confers to such sources an even greater single-photon source
character.  As can be expected, correlations in time are affected by
dephasing as well as by the detuning between the two-level system, the
cavity and the laser. Such a characteristic profile is shown as filled
blue lines in Fig.~\ref{fig:MonApr16210113BST2018}, to which we
compare the cases treated in this text. Panels~(a) and~(b) show the
effect of detuning (without dephasing) and Panel~(c) shows the effect
of dephasing (without detuning). In~(a) the detector is resonant with
the two-level system, and both are detuned from the driving laser,
while in~(b), the detector is resonant with the laser, and both are
detuned from the emitter. In green lines are shown the correlations
for the given parameters featured in inset while red lines show the
result without the laser correction, i.e., with~$\mathcal{F}=0$.

In Fig.~\ref{fig:MonApr16210113BST2018}(a), when the driving laser is
detuned, the zero delay correlations can be compensated exactly, but
the antibunching plateau is lost or at least strongly reduced, and the
correlations display fast oscillations between antibunching and
bunching. Conversely, when the detector is resonant to the laser,
panel~(b), the plateau of antibunching is still present, albeit for a
shorter time, and the correlations do not display any
oscillations. This means that, although the configuration with the
detector in resonance to the emitter could seem the more natural or
appealing one, the fast oscillations in its delay correlations makes
it potentially problematic, and if it would turn out to be impossible
to bring the driving laser in resonance with the emitter, then the
detection should be done at the frequency of the laser. It is also
shown how the laser-correction makes a huge improvement on the
antibunching as compared to the standard case which features almost no
antibunching. However, a very large detuning between the two-level
system and the laser washes out the plateau and the correlations
become simply
$g_a^{(2)}(\tau) = \left( 1- e^{-\Gamma \tau/2}\right)^2$, which
corresponds to the correlations of a two-level system of
linewidth~$\Gamma$ driven coherently in the regime of low
excitation. In Fig.~\ref{fig:MonApr16210113BST2018}(c), it is shown
how the zero-delay correlation of the dephased two-level system
\emph{cannot} be compensated exactly in presence of dephasing, as
already stated, but otherwise suffers little in term of its plateau or
coherence time. More importantly, it remains largely improved as
compared to the case without the interference, with a value
of~$g_a^{(2)}(0)\approx 0.05$ for a dephasing rate of 10\% the emitter
decay rate, whereas it is only~$g_a^{(2)}(0)\approx 0.73$ without the
laser correction. Furthermore, the measured linewidth remains well
below the natural (but broadened by the dephasing) linewidth of the
two-level system.

\section{Conclusions}
\label{sec:FriApr20161021BST2018}

We have discussed our scheme to produce joint antibunching and
subnatural linewidth emission~\cite{arXiv_lopezcarreno18a} in the
presence of dephasing and of a detuning from the emitter with either
the driving laser and/or the detector. Not only does this scheme makes
these properties hold simultaneously, it also produces perfect
antibunching to first order in the driving (that is, $g_a^{(2)}(0)=0$)
which is otherwise possible only by integrating all frequencies, and
features a plateau in the time-delayed photon correlations, making
such a single-photon source more effectively suppressing coincidences.
We have also shown that different conditions from the interfering
laser produce instead perfect superbunching (that is,
$g_a^{(2)}(0)=\infty$ to first order in the driving).

We have shown how, in presence of detuning alone, perfect antibunching
can always be enforced by the interference, but result in strong time
oscillations of the statistics when the laser is detuned and the
detector is resonant to the laser. On the other hand, no such
oscillations are featured when detecting at the laser frequency, hence
favouring this configuration. In presence of dephasing, antibunching
is spoilt and only finite values can be obtained. However, the
improvement as compared to the non-corrected single-photon source is
also dramatic, making this scheme of value in this case as well.

\section*{Acknowledgments}


Funding from the joint Russian-Greek project RFMEFI61617X0085
supported by the Ministry of Science and Education of the Russian
Federation, the Universidad Aut\'onoma de Madrid under contract
FPI-UAM 2016, the Spanish MINECO under contract FIS2015-64951-R
(CLAQUE) \& the RyC program is gratefully acknowledged.

\section*{References}

\bibliographystyle{naturemag}
\bibliography{Sci,books,arXiv} 



 \pagebreak
\appendix
\setcounter{equation}{0}
\setcounter{figure}{0}

\section{Wavefunction approximation method at vanishing pumping
  regime} 
\label{app:3}
In the context of this work, the wavefunction
approximations~\cite{visser95a} consist of assuming that the state of
the system composed by two fields, with annihilation operators~$\xi$
and~$c$ following either pseudo-spin or bosonic algebra, can be
approximated by a pure state, which reads in the Fock state basis
\begin{equation}
  \label{eq:WedFeb28112316CET2018}
\ket{\psi} = \sum_{n,m} \mathcal{C}_{nm} \ket{n}_c\ket{m}_\xi \equiv
\sum_{n,m} 
\mathcal{C}_{nm} \ket{n\,,m} \,,
\end{equation}
where~$\mathcal{C}_{nm}$ are the probability amplitude of having~$n$
photons in the field of operator~$\xi$ and~$m$ photons in the field of
operator~$c$. The summation extends over the boundaries of the
respective spaces, which is~1 for a two-level system and~$\infty$ for
a bosonic one, which in practice is truncated to~$N$. Since the
dynamics of the system is given by the master equation
\begin{equation}
\label{eq:WedFeb28114318CET2018}
\partial_t \rho = i[\rho,H] + \sum_k (\tilde \Gamma_k/2)
\mathcal{L}_{j_k}\rho\,,
\end{equation}
where~$H$ is the Hamiltonian of the system and assuming the
dissipation in the form of ``jump operators''~$j_k$ at
rates~$\tilde\Gamma_k$, the dynamics of the wavefunction is given by
Sch\"odinger equation
\begin{equation}
  \label{eq:WedFeb28113458CET2018}
  \partial_t \ket{\psi} = - i H_\mathrm{eff}\ket{\psi}\,,
\end{equation}
where~$H_\mathrm{eff}$ is a non-hermitian Hamitonian constructed as~$
H_\mathrm{eff}=H-i \sum_k \tilde\Gamma_k\, \ud{j_k} j_k$, and the
coefficients evolve as
\begin{equation}
\label{eq:coeffeqs}
\partial_t \, \mathcal{C}_{nm} = -i  \sum_{p,q} \bra{n\,,m}H_\mathrm{eff}
\ket{p\,,q} \mathcal{C}_{pq}\,. 
\end{equation}

In our particular case, in which we describe the excitation of a
sensor (an harmonic oscillator) by the emission of a two-level system,
which in turn is driven in the Heitler regime by a laser, the
Hamiltonian is the one given in Eq.~(\ref{eq:TueMar13171915GMT2018})
of the main text:
\begin{equation}
  \label{eq:MonMar5160049CET2018}\fl
  H = \Delta_\sigma \ud{\sigma}\sigma + \Delta_a \ud{a}a + \Omega_\sigma (\ud{\sigma} + \sigma) + gt
  (\ud{\sigma}a+\ud{a}\sigma) -i r |\beta| (\ud{a} e^{i\phi}  -
  a e^{-i\phi})\,.
\end{equation}
Here the two-level system is driven with intensity~$\Omega_\sigma$ and
is coupled to the cavity with strength~$g$, the cavity is also driven
by a field~$\beta e^{i\phi}$ and the detuning between the two-level
system (resp. sensor) and the driving laser is given
by~$\Delta_\sigma$ (resp.~$\Delta_a$). These fields are attenuated by
the transmission~$t$ and reflection~$r$ coefficients of the beam
splitter in which they interfere. Considering that the two-level
system and the cavity have decay rates~$\gamma_\sigma$ and~$\Gamma$,
respectively, the effective Hamiltonian that describes the dynamics in
the wavefunction approximation reads\footnote{The dephasing of the
  two-level system enters the description as an extra Lindblad term in
  the master equation:
  $(\gamma_\phi/2) \mathcal{L}_{\ud{\sigma}\sigma} \rho$,
  where~$\gamma_\phi$ is the rate of dephasing. However, the effect of
  this term is the decoherence of the state of the two-level system,
  which affects only the off-diagonal elements of the \emph{density
    matrix} of the two-level system, and thus cannot be described
  through a wavefunction approximation.}
\begin{equation}
  \label{eq:MonMar5160912CET2018}
  H_\mathrm{eff} = H -\frac{i}{2}\left ( \gamma_\sigma
    \ud{\sigma}\sigma + \Gamma \ud{a}a \right)\,,
\end{equation}
where~$H$ is the Hamiltonian in
Eq.~(\ref{eq:MonMar5160049CET2018}). Replacing the effective
Hamiltonian in Eq.~(\ref{eq:MonMar5160912CET2018}) in the expression
in Eq.~(\ref{eq:WedFeb28113458CET2018}), we obtain the differential
equations for the coefficients of interest:
  \begin{eqnarray}
    \label{eq:MonMar5161258CET2018a}\fl
    i \partial_t \mathcal{C}_{01} &= \Omega_\sigma + gt
                                    \mathcal{C}_{10} +
                                    +ir |\beta|
                                    e^{-i\phi}  
                                    \mathcal{C}_{11}+\left( \Delta_\sigma-
                                    i\frac{\gamma_\sigma}{2} \right)
                                    \mathcal{C}_{01}\,,\\
    \label{eq:MonMar5161258CET2018b}\fl
    i \partial_t \mathcal{C}_{10} &= -ir
                                    |\beta|
                                    e^{i\phi}
                                    + \Omega_\sigma 
                                    \mathcal{C}_{11}+ 
                                    gt \mathcal{C}_{01} +  \sqrt{2}
                                    ir|\beta| 
                                    e^{-i\phi} 
                                    \mathcal{C}_{20}+\left(\Delta_a -
                                    i\frac{\Gamma}{2} \right)
                                    \mathcal{C}_{10}\,, \\
    \label{eq:MonMar5161258CET2018c}\fl
    i \partial_t \mathcal{C}_{11} &= \Omega_\sigma \mathcal{C}_{10} -
                                    ir|\beta| e^{i\phi}
                                    \mathcal{C}_{01}+\sqrt{2}
                                    gt\mathcal{C}_{20}+
                                    \left(\Delta_\sigma +
                                    \Delta_\sigma
                                    -i\frac{\gamma_\sigma+\Gamma}{2}\right)
                                    \mathcal{C}_{11}\,,\\
    \label{eq:MonMar5161258CET2018d}\fl
    i \partial_t \mathcal{C}_{20} & = \sqrt{2}gt\mathcal{C}_{11} -
                                    \sqrt{2} i r|\beta|e^{i
                                    \phi}\mathcal{C}_{10}
                                    +2\left(\Delta_a -
                                    i\frac{\Gamma}{2} \right)
                                    \mathcal{C}_{20}\,, 
  \end{eqnarray}
where we have assumed that the driving to the two-level system is low
enough so that the states with three or more excitations can be safely
neglected, and that the driving laser is resonant to both the
two-level system and the sensor. Assuming that the coherent field that
drives the sensor can be written as a fraction of the field that
drives the two-level system, as in
Eq.~(\ref{eq:SatFeb24130524CET2018}), and to leading order in the
coupling and the driving intensity of the two-level system, the
solution to
Eqs.~(\ref{eq:MonMar5161258CET2018a}-\ref{eq:MonMar5161258CET2018d})
is
\begin{eqnarray}\fl
  \label{eq:MonMar5165614CET2018a}
  \mathcal{C}_{01} &=&  -\frac{2i\Omega_\sigma}{\gamma_\sigma+
                         2i\Delta_\sigma}\,,\\\fl  
  \label{eq:MonMar5165614CET2018b}
  \mathcal{C}_{10} &=& -\frac{2gt\Omega_\sigma [2\gamma_\sigma +
                       (\gamma_\sigma +2i\Delta_\sigma) \mathcal{F}e^{i\phi}]}
                       {\gamma_\sigma (\Gamma + 2i\Delta_a)
                       (\gamma_\sigma + 2i\Delta_\sigma)}\,,\\\fl
  \label{eq:MonMar5165614CET2018c}
    \mathcal{C}_{11} &=& \frac{4igt\Omega_\sigma^2 [2\gamma_\sigma
                         +(\gamma_+ + 2i\Delta_+)
                         \mathcal{F}e^{i\phi}]}{ \gamma_\sigma
                         (\Gamma + 2i\Delta_a) 
                       (\gamma_\sigma + 2i\Delta_\sigma) (\gamma_+ +
                         2i\Delta_+)}\,,\\\fl
  \label{eq:MonMar5165614CET2018d}
    \mathcal{C}_{20}&=& \frac{2\sqrt{2} g^2 t^2 \Omega_\sigma^2 \{
                        4\gamma_\sigma^2 + (\gamma_+ +2i\Delta_+)
                        e^{i\phi} \mathcal{F} [4
                        \gamma_\sigma  +
                        e^{i\phi} \mathcal{F} ( \gamma_\sigma + 2i
                        \Delta_\sigma) ] \} }{\gamma_\sigma^2 (\Gamma +
                        2i\Delta_a)^2  
                       (\gamma_\sigma + 2i\Delta_\sigma) (\gamma_+ + 
                         2i\Delta_+)}\,.
\end{eqnarray}
The population of both the two-level system and the cavity, as well
as~$\g{2}_a$ can be obtained from the coefficients in
Eqs.~(\ref{eq:MonMar5165614CET2018a}-\ref{eq:MonMar5165614CET2018d})
as~$ n_a = |\mathcal{C}_{10}|^2$,
$\mean{n_\sigma}= |\mathcal{C}_{01}|^2$
and~$\g{2}_a = 2|\mathcal{C}_{20}|^2/ |\mathcal{C}_{10}|^4$,
respectively. The cancellation of the coefficient~$\mathcal{C}_{20}$,
and therefore of~$\g{2}_a$, yields the condition on the attenuation
factor
\begin{equation}
  \label{eq:TueMar6111957CET2018}
  \mathcal{F} =  -\frac{2\gamma_\sigma e^{-i\phi}}{\gamma_\sigma +2
    i\Delta_\sigma} \left ( 1 \pm \sqrt{
      \frac{\Gamma+2i\Delta_a}{\Gamma_+ +2i\Delta_+} } \right)\,,
\end{equation}
in agreement with Eq.~(\ref{eq:MonMar12172809GMT2018}) of the main
text.

\end{document}